\documentclass{article}
\usepackage[english]{babel}
\usepackage{cite} 
\usepackage{fancyhdr}
\usepackage{lipsum} 

\usepackage{graphicx,color,bm}

\usepackage[letterpaper,top=2cm,bottom=2cm,left=3cm,right=3cm,marginparwidth=1.75cm]{geometry}

\usepackage{indentfirst}
\usepackage{amsmath}
\usepackage{graphicx}
\usepackage{slashed}
\usepackage[colorlinks=true, allcolors=blue]{hyperref}

\fancypagestyle{firstpage}{
    \fancyhf{} 
    \fancyhead[L]{\textit{To be published in Chinese Physics C}} 
    \fancyhead[R]{} 
    \fancyfoot[C]{\thepage} 
}

\newcommand{\keywords}[1]{\noindent\textbf{keywords:} #1}

\makeatletter
\newcommand{\affiliation}[1]{\gdef\@affiliation{#1}}
\def\@maketitle{%
  \newpage
  \null
  \vskip 2em%
  \begin{center}%
  \let \footnote \thanks
    {\LARGE \@title \par}%
    \vskip 1.5em%
    {\large
      \lineskip.5em%
      \begin{tabular}[t]{c}%
        \@author
      \end{tabular}\par}%
    \vskip 1em%
    {\large \@affiliation \par}%
  \end{center}%
  \par
  \vskip 1.5em}
\makeatother

\title{The asymmetry $A_{LL}^{\cos2\phi}$ in the polarized proton-proton Drell-Yan process within TMD factorization}

\author{Hui Li$^{1,a}$, Ting Hu$^{1}$, Liang-Liang Liu$^{1}$,\\ Xiao-Yu Wang$^{2,b}$, Zhun Lu$^{3,c}$ }
\affiliation{$^{1}$\quad School of Physics and Information Engineering, Shanxi Normal University, Taiyuan 030031, China\\
$^{2}$\quad School of Physics, Zhengzhou University, Zhengzhou 450001, China\\
            $^{3}$\quad School of Physics, Southeast University, Nanjing 211189, China}
\begin{document}
\maketitle
\thispagestyle{firstpage}
\let\thefootnote\relax
\footnote{
$^{\ast}$Hui Li supported by Fundamental Research Program of Shanxi Province (No. 202203021222224) and Natural Science Foundation of Shanxi Normal University (No. JCYJ2023021). Xiao-Yu Wang supported by the Natural Science Foundation of Henan Province under grant
No. 242300421377, 232300421140.\\
$^{a}$E-mail: lihui@sxnu.edu.cn\\
$^{b}$E-mail: xiaoyuwang@zzu.edu.cn\\ 
$^{c}$E-mail: zhunlu@seu.edu.cn}
\begin{abstract}
We study the $\cos2\phi$ azimuthal asymmetry in doubly longitudinally polarized proton-proton Drell-Yan collisions within the transverse momentum dependent factorization framework. The asymmetry arises from the convolution of the longitudinal transversity distribution $h_{1L}^{\perp}$ for both protons. Using the Bacchetta-Delcarro-Pisano-Radici-Signori parametrization for the nonperturbative Sudakov form factor and the Wandzura-Wilczek approximation for the collinear $h_{1L}^{\perp}$, we predict the double spin asymmetry $A_{LL}^{\cos2\phi}$ at RHIC and NICA kinematics. Our results demonstrate sensitivity to sea quark distributions, with the asymmetry reaching up to $25\%$ for maximal sea quark contributions. These predictions highlight the potential of polarized Drell-Yan measurements to probe sea quark dynamics and advance our understanding of nucleon structure.
\end{abstract}

\keywords{Transverse momentum dependent factorization; the longitudinal transversity distribution; Relativistic Heavy Ion Collider; Nuclotron-based Ion Collider Facility}

\section{Introduction}

The three-dimensional tomography of nucleons represents a fundamental frontier in quantum chromodynamics (QCD). While collinear parton distribution functions (PDFs) provide longitudinal momentum information, transverse momentum dependent parton distribution functions (TMDPDFs)~\cite{Mulders:1995dh,Barone:2001sp,Bacchetta:2006tn,Barone:2010zz} offer crucial insights into the transverse momentum ($\bm{k}_\perp$) structure, enabling a complete momentum-space description of nucleon constituents. Formally, the quark-quark correlator for a nucleon with momentum $P$ and spin vector $S$ can be expressed as:
\begin{equation}
\Phi_{ij}(x,\bm{k}_\perp) = \int\frac{d\xi^-d^2\bm{\xi}_\perp}{(2\pi)^3} e^{ik\cdot\xi} \langle P,S|\bar{\psi}_j(0)\mathcal{U}_{[0,\xi]}\psi_i(\xi)|P,S\rangle\big|_{\xi^+=0},
\end{equation}
where $\mathcal{U}_{[0,\xi]}$ denotes the gauge link ensuring color gauge invariance. Through Lorentz structure decomposition~\cite{Mulders:1995dh,Boer:1997nt}, 
eight leading-twist TMDPDFs emerge, characterized by distinct spin-momentum correlations.
Three TMDPDFs, the unpolarized distribution $f_1(x,\boldsymbol{k}_{\perp})$, the helicity distribution $g_1(x,\boldsymbol{k}_{\perp})$ and transversity distribution $h_1(x,\boldsymbol{k}_{\perp})$, reduce to the corresponding collinear PDFs when the transverse momentum is integrated over. 
The remaining five TMDPDFs, depending on the spin of the parent proton or the quarks, encode genuine transverse momentum effects. 
                                                                                                                                                                                                                                                                                                                                                                                                                                                                                                                                                                                                                                                                                                                                                                                                                          
Among the TMDPDFs, the distribution $h_{1L}^{\perp}$ is arguably the least explored. It quantifies the correlation between a longitudinally polarized nucleon and transversely polarized quarks through the matrix element:
\begin{equation}
h_{1L}^\perp \propto \epsilon_T^{\rho\sigma}k_{T\rho}S_{L\sigma} \langle \bar{\psi}\gamma^+ \psi \rangle,
\end{equation}
where $\epsilon_T^{\rho\sigma}$ is the transverse Levi-Civita tensor and $S_L$ the longitudinal spin vector.
For this reason, $h_{1L}^{\perp}$ is usually called as longitudinal transversity (abbreviated as longi-transversity).
Another peculiarity of $h_{1L}^{\perp }$ is its chiral-odd nature, necessitating partner chiral-odd functions (e.g., Collins fragmentation function~\cite{Kotzinian:1994dv,Kotzinian:1997wt}) for experimental accessibility via single spin asymmetry $A_{UL}^{\sin2\phi}$ in semi-inclusive deep inelastic scattering (SIDIS). 
Predictions for this asymmetry, as reported in Refs.~\cite{HERMES:1999ryv,HERMES:2002buj,CLAS:2010fns,
Lu:2011pt,Zhu:2011zza,Boffi:2009sh,Ma:2000ip,Ma:2001ie,Li:2021mmi}, have been found to be approximately at the few-percent level, though systematic uncertainties remain significant.

It is also possible to probe the distribution $ h_{1L}^{\perp} $ through the Drell-Yan process. As demonstrated in Refs.~\cite{Arnold:2008kf,Zhu:2011ym}, the combination of $ h_{1L}^{\perp} $ contributions from both colliding protons can generate a distinctive $\cos2\phi$ azimuthal asymmetry in doubly longitudinally polarized Drell-Yan process, where $\phi$ denotes the azimuthal angle of the dilepton system with respect to the hadronic plane. Typically, the double spin asymmetry $A_{LL}^{\cos2\phi}$ in the proton-proton Drell-Yan process is expected to be small, as it is proportional to the product of the polarized quark distribution with the antiquark one. However the polarized $ p^{\to}p^{\to} $ Drell-Yan process provides unique sensitivity to sea quark dynamics within the proton. Notably, despite its theoretical significance, no systematic phenomenological investigation of this asymmetry has been reported in existing literature. This motivates our primary objective: a comprehensive analysis of $ A_{LL}^{\cos2\phi} $ in the polarized Drell-Yan process.

In this work, we present a detailed phenomenological study of the double longitudinal spin asymmetry $ A_{LL}^{\cos2\phi} $ in the polarized $pp$ Drell-Yan process within the TMD factorization framework~\cite{Collins:1981uk,Collins:1984kg,Collins:2011zzd,Aybat:2011zv,Collins:2012uy,
Echevarria:2012js,Pitonyak:2013dsu,Echevarria:2014xaa,Kang:2015msa,
Bacchetta:2017gcc,Wang:2017zym,Wang:2018pmx,Li:2019uhj}. 
Our predictions are computed for kinematic configurations accessible at both the Relativistic Heavy Ion Collider (RHIC) and the Nuclotron-based Ion Collider Facility (NICA) \cite{Lu:2011cw}. 
Utilizing the TMD factorization, we provide the expressions of the spin-dependent cross section for $ p^{\to}p^{\to} \to l^+l^-X $ alongside the unpolarized one, with the asymmetry defined as the ratio of these two cross sections. 

Over the past two decades, TMD factorization has emerged as a powerful tool for exploring the three-dimensional structure of the nucleon, finding widespread application in various high-energy processes. The TMD factorization theorem expresses the differential cross section in the small transverse momentum region of the lepton (\( \boldsymbol{q}_{\perp} \ll Q \), where \( Q \) is the invariant mass of the dilepton pair) as a convolution of two contributions: one representing hard scattering factors at short distances, and the other accounting for coherent long-distance interactions, described in terms of well-defined TMDs. In our analysis, the \( \cos2\phi \) asymmetry arises from the convolution of \( h_{1L}^{\perp} \) (for both protons) with the hard scattering factors. Additionally, the TMD formalism encodes the evolution of TMDs, governed by the Collins-Soper equation~\cite{Collins:1981uk,Collins:1984kg,Collins:2011zzd,Idilbi:2004vb}. The solution to this equation is typically expressed as an exponential form of the Sudakov-like form factor~\cite{Aybat:2011zv,Collins:1984kg,Collins:2011zzd,Collins:1999dz}, which describes the transformation of TMDs from an initial scale to another scale. The Sudakov factor decomposes into perturbative and nonperturbative components: while the perturbative part admits a universal operator product expansion, the nonperturbative contribution requires phenomenological parametrization constrained by experimental data. 
Several nonperturbative components of the Sudakov form factor for TMDs have been extracted from experimental data in the literature~\cite{Aybat:2011zv,Echevarria:2012js,Echevarria:2014xaa,Bacchetta:2017gcc,Collins:1984kg,
Collins:2011zzd,Kang:2011mr,Echevarria:2014rua,Landry:2002ix,Davies:1984sp,Ellis:1997sc,
Aybat:2011ge,Sun:2014dqm,Collins:2014jpa,Nadolsky:1999kb,Aidala:2014hva,Konychev:2005iy}.
For our analysis, we adopt the BDPRS parametrization \cite{Bacchetta:2017gcc} for the $ h_{1L}^{\perp} $-associated nonperturbative Sudakov factor, building upon established extractions from deep inelastic scattering and Drell-Yan data \cite{Aybat:2011zv,Echevarria:2012js,Echevarria:2014xaa,Bacchetta:2017gcc}.

The remaining content of the paper is organized as follows. In Sec.~\ref{Sec.formalism} we present the formalism of the asymmetry $A_{LL}^{\cos2\phi}$ in the process $p^{\to }p^{\to}\rightarrow l^{+}l^{-}X$ within the TMD factorization. In Sec.~\ref{Sec.evolution}, we investigate the evolution effect of the distribution $h_{1L}^{\perp }$. Particularly, we discuss the parametrization of the non-perturbative Sudakov form factors associated with the $h_{1L}^{\perp }$ in details. 
In Sec.~\ref{Sec.numerical}, we present the phenomenological predictions for the asymmetry $A_{LL}^{\cos2\phi}$  at RHIC and NICA. We conclude our paper in  Sec.~\ref{Sec.conclusion}.

\section{Formalism of the asymmetry $A_{LL}^{\cos2\phi}$ in the doubly  polarized proton-proton Drell-Yan process}
\label{Sec.formalism}

In this section, we present the theoretical framework for calculating the double longitudinal spin asymmetry \( A_{LL}^{\cos2\phi} \) in polarized proton-proton Drell-Yan collisions within the TMD factorization formalism~\cite{Collins:2011zzd}, incorporating the scale evolution of TMDs. 
The asymmetry originates from the convolution of the \( h_{1L}^{\perp} \) distributions of both colliding protons at leading twist. The specific process under study is described as follows \cite{Arnold:2008kf}:
\begin{align}
p^{\rightarrow }\left(P_{1}, S_1\right)+p^{\rightarrow }\left(P_{2}, S_2\right) \rightarrow \gamma^{*}(q)+ X\rightarrow l^{+}(\ell)+l^{-}\left(\ell^{\prime}\right)+X.
\label{eq:DY-process}
\end{align}
where $P_{1/2}, S_{1/2} $ denote the four-momenta and spin vectors of the incoming protons, respectively. The virtual photon momentum $ q $ is time-like, distinguishing this process from semi-inclusive deep inelastic scattering (SIDIS). Here, $ Q^{2} = q^{2} $ represents the invariant mass squared of the lepton pair, and the notation $\rightarrow $ indicates longitudinal polarization of the protons.

The following kinematic variables are usually introduced to characterize the experimental observables,
\begin{align}
&s=\left(P_{1}+P_{2}\right)^{2}, ~~\quad x_{1}=\frac{Q^{2}}{2 P_{1} \cdot q},~~ \quad x_{2}=\frac{Q^{2}}{2 P_{2} \cdot q}, \nonumber\\&
x_{F}=2 q_L / s=x_{1}-x_{2}, \quad \tau=Q^{2} / s=x_{1} x_{2}, \quad y=\frac{1}{2} \ln \frac{q^{+}}{q^{-}}=\frac{1}{2} \ln \frac{x_{1}}{x_{2}},&
\label{eq:variable1}
\end{align}
where $s$ represents the total center of mass energy squared. The variables $x_{1/2}$ denotes the longitudinal momentum fraction. $q_{L}$ stands for the longitudinal momentum of the virtual photon, while $x_{F}$ is the Feynman $x$ variable.  The variable $y$ corresponds to the rapidity of the dilepton. Furthermore, $x_{1/2}$ can be expressed as functions of $x_{F}$, $\tau $ and $y$, $\tau$, respectively, as follows
\begin{align}
x_{1/2}=\frac{\pm x_{F}+\sqrt{x_{F}^{2}+4\tau }  }{2}  ,~~~
x_{1/2}=\sqrt{\tau  }e^{\pm y} .
\label{eq:variable2}
\end{align}

In Drell-Yan processes, when the transverse momentum $\boldsymbol{q}_{\perp}$ of the dilepton pair is measured, the TMD factorization framework becomes applicable in the kinematic regime $\boldsymbol{q}_{\perp} \ll Q$ \cite{Aybat:2011zv,Collins:2012uy,Echevarria:2012js,Pitonyak:2013dsu,Echevarria:2014xaa,Kang:2015msa,Bacchetta:2017gcc,Wang:2017zym,Wang:2018pmx,Li:2019uhj}. Within this framework, the differential cross section at leading twist can be expressed as \cite{Zhu:2011ym}:
\begin{align}
&\frac{d \sigma}{d x_{1} d x_{2} d^{2} \boldsymbol{q}_{\perp}d\Omega} \nonumber\\&
=\frac{\alpha_{e m}^{2}}{4 Q^{2}}\left\{\left(1+\cos^{2} \theta\right) F_{U U}+S_{1 L} S_{2 L} \sin ^{2} \theta \cos 2 \phi F_{L L}^{\cos 2 \phi}\right. \nonumber\\&
\quad+\left|\boldsymbol{S}_{1 T}\right| S_{2 L}\left(1+\cos ^{2} \theta\right) \cos \phi_{1} F_{T L}^{\cos \phi_{1}} \nonumber\\&
\quad+S_{1 L}\left|\boldsymbol{S}_{2 T}\right| \sin ^{2} \theta\left[\cos \left(2 \phi+\phi_{2}\right) F_{L T}^{\cos \left(2 \phi+\phi_{2}\right)}\right. \nonumber\\&
\left.\quad+\cos \left(2 \phi-\phi_{2}\right) F_{L T}^{\cos \left(2 \phi-\phi_{2}\right)}\right] \nonumber\\&
\quad+\left|\boldsymbol{S}_{1 T}\right|\left|\boldsymbol{S}_{2 T}\right|\left(1+\cos ^{2} \theta\right)\left[\left(\cos \left(\phi_{1}+\phi_{2}\right) F_{T T}^{\cos \left(\phi_{1}+\phi_{2}\right)}\right.\right. \nonumber\\&
\left.\left.\left.\quad+\cos \left(\phi_{1}-\phi_{2}\right) F_{T T}^{\cos \left(\phi_{1}-\phi_{2}\right)}\right)\right]+\cdots\right\},&
\label{eq:cross section}
\end{align} 
where $\alpha_{\text{em}}$ is the fine-structure constant, $\phi_{1/2}$ denotes the azimuthal angle of the transverse spin vector $\boldsymbol{S}_{(1/2)T}$ with respect to the lepton plane, and $\phi$ and $\theta$ represent the azimuthal and polar angles of the lepton momentum in the Collins-Soper frame, respectively. The solid angle $\Omega$ specifies the orientation of the dilepton system.

Furthermore, $F_{P}^{f\left [ \phi ,\phi _{1/2}  \right ] } $ denotes the structure
functions with a specific modulation $f\left [ \phi ,\phi _{1/2}  \right ]$, with $P=UU,~P=LL,~P=TL,~P=LT,~P=TT$ denoting the polarization of the incoming protons (U for unpolarized, T for transversely polarized, L for longitudinally polarized). In this work, the relevant structure functions are $F_{UU}^{1}$, representing the unpolarized structure function, and $F_{LL}^{\cos 2 \phi}$, corresponding to the spin-dependent structure function. The ratio of these two structure functions 
defines the $\cos 2\phi$ azimuthal asymmetry 
\begin{align}
A_{LL}^{\cos2\phi } =\frac{F_{LL}^{\cos2\phi }}{F_{UU}},
\label{eq:asymmetry}
\end{align}
which could be measured in double-longitudinally polarized Drell-Yan processes.

The spin-averaged structure function $F_{UU}$ can be expressed as the convolution of the unpolarized distribution functions from each proton~\cite{Zhu:2011ym,Arnold:2008kf}
\begin{align}
 F_{UU}= \mathcal{C} \left [ f_{1}^{q/p} f_{1}^{\bar{q}/p} \right ],
\label{eq:SF-UU}
\end{align}
while the spin-dependent structure function $F^{\cos2\phi}_{LL}$ is expressed as the convolution of the longi-transversity distributions~\cite{Zhu:2011ym,Arnold:2008kf}
\begin{align}
F_{L L}^{\cos 2 \phi}=\mathcal{C}\left[\frac{2\left(\boldsymbol{\hat{h}} \cdot \boldsymbol{k}_{1\perp}\right)\left(\boldsymbol{\hat{h}} \cdot \boldsymbol{k}_{2\perp}\right)-\boldsymbol{k}_{1\perp} \cdot \boldsymbol{k}_{2\perp}}{M^2} h_{1 L}^{\perp,q/p} h_{1 L}^{\perp,\bar{q}/p}\right],
\label{eq:SF-LL}
\end{align}
where the unit vector $\boldsymbol{h}$ is defined as $\boldsymbol{h}\equiv\boldsymbol{q}_{\perp} /|\boldsymbol{q}_{\perp}|$, and $M$ represents the mass of the proton~\cite{Boer:1999mm,Arnold:2008kf}. $\boldsymbol{k}_{1\perp}$ and $\boldsymbol{k}_{2\perp}$ denote the transverse momentum of the quark and antiquark in the incoming protons, respectively. The convolution of TMDPDFs in the transverse momentum space is defined as follows~\cite{Arnold:2008kf}
\begin{align}
&\mathcal{C}\left[w\left(\boldsymbol{k}_{1\perp}, \boldsymbol{k}_{2\perp}\right) f_{1} f_{2}\right] \nonumber\\&
=\frac{1}{N_{c}} \sum_{q} e_{q}^{2} \int d^{2} \boldsymbol{k}_{1\perp} d^{2} \boldsymbol{k}_{2\perp} \delta^{(2)}\left(\boldsymbol{q}_{\perp}-\boldsymbol{k}_{1\perp}-\boldsymbol{k}_{2\perp}\right) \nonumber\\&
\times\omega\left(\boldsymbol{k}_{1\perp}, \boldsymbol{k}_{2\perp}\right) f_{1}^{q}\left(x_{1}, \boldsymbol{k}_{1\perp}^{2}\right) f_{2}^{\bar{q}}\left(x_{2}, \boldsymbol{k}_{2\perp}^{2}\right),
\label{eq:convolution}
\end{align}
with $N_{c} = 3$ representing the number of colors, $\boldsymbol{q}_{\perp}$ denoting the transverse momenta of the lepton pair. Finally, $\omega(\boldsymbol{k}_{1\perp},\boldsymbol{k}_{2\perp})$ is a function of $\boldsymbol{k}_{1\perp}$ and $\boldsymbol{k}_{2\perp}$. 

Generally, a more feasible approach to studying the structure function is the $b_{\perp}$-space framework, where the convolution of TMDs can be simplified into product of $b_{\perp}$-dependent TMDs. Subsequently, the physical observables can be derived via Fourier transformation from the $b_{\perp}$-space to the $k_{\perp}$-space.

By using the Fourier transformation of the $\delta$ function
\begin{align}
\delta ^{2}(\boldsymbol{q}_{\perp}-\boldsymbol{k}_{1\perp}-\boldsymbol{k}_{2\perp})=\frac{1}{(2\pi)^{2}}\int d^{2} \textbf{\textit{b}}_{\perp}e^{i\textit{\textbf{b}}_{\perp}(\boldsymbol{q}_{\perp}-\boldsymbol{k}_{1\perp}-\boldsymbol{k}_{2\perp})},
\label{eq:Fourier transform}
\end{align}
one can obtain the spin-dependent structure function $F_{LL}^{cos2\phi }$, which is given by
\begin{align}
&F_{LL}^{cos2\phi }=\frac{1}{N_{c}} \sum_{q} e_{q}^{2} \int d^{2} \boldsymbol{k}_{1\perp} d^{2} \boldsymbol{k}_{2\perp} \int \frac{d^{2} \boldsymbol{b}_{\perp}}{(2 \pi)^{2}} e^{i\left(\boldsymbol{q}_{\perp}-\boldsymbol{k}_{1\perp}-\boldsymbol{k}_{2\perp}\right) \cdot \boldsymbol{b}_{\perp}}  \nonumber \\
&\times \left [ \frac{2(\boldsymbol{h}\cdot \boldsymbol{k}_{1\perp})(\boldsymbol{h}\cdot \boldsymbol{k}_{2\perp})-(\boldsymbol{k}_{1\perp} \cdot \boldsymbol{k}_{2\perp})}{M^{2}} \right ]  h^{\perp , q / p}_{1L}\left(x_{1}, \boldsymbol{k}_{1 \perp}^{2},Q\right)  h^{\perp , \bar{q} /p}_{1L}\left(x_{2}, \boldsymbol{k}_{2 \perp}^{2},Q\right)\nonumber\\
&=\frac{1}{N_{c}} \sum_{q}^{}e_{q}^{2}\int \frac{d^{2}\boldsymbol{b}_{\perp}}{(2\pi)^{2}}e^{i\boldsymbol{q_{\perp}}\cdot \boldsymbol{b}_{\perp}}\left ( 2\boldsymbol{h}_{\alpha }\cdot \boldsymbol{h}_{\beta}-g^{\alpha\beta} \right ) \cdot\tilde{h}_{1L}^{\perp~q/p,\alpha}(x_{1},\boldsymbol{b}_{\perp},Q) \tilde{h}_{1L}^{\perp~\bar{q}/p,\beta}(x_{2},\boldsymbol{b}_{\perp},Q).
\label{eq:F-LL}
\end{align} 
Here, the tilde terms represent the ones in the $b_{\perp}$-space. 
The longi-transversity of the incoming protons in the $b_{\perp}$-space is defined as follows
\begin{align}
\tilde{h}_{1L}^{\perp q/p,\alpha}(x_1,\boldsymbol{b}_{\perp};Q)=\int d^{2}\boldsymbol{k}_{1\perp}e^{-i\boldsymbol{k}_{1\perp}\cdot \boldsymbol{b}_{\perp}}\frac{\boldsymbol{k}_{1\perp}^{\alpha}}{M}h_{1L}^{\perp q/p} (x_1,\boldsymbol{k}_{1\perp}^2;Q),
\label{eq:h1L-perp-alpha}
\end{align}
\begin{align}
\tilde{h}_{1L}^{\perp \bar{q}/p,\beta}(x_2,\boldsymbol{b}_{\perp};Q)=\int d^{2}\boldsymbol{k}_{2\perp}e^{-i\boldsymbol{k}_{2\perp}\cdot \boldsymbol{b}_{\perp}}\frac{\boldsymbol{k}_{2\perp}^{\beta}}{M}h_{1L}^{\perp \bar{q}/p} (x_2,\boldsymbol{k}_{2\perp}^2;Q).
\label{eq:h1L-perp-beta}
\end{align}

Similarly, the spin-averaged structure function $F_{UU}$ can be expressed as
\begin{align}
&F_{UU}=\frac{1}{N_{c}} \sum_{q} e_{q}^{2} \int d^{2} \boldsymbol{k}_{1\perp} d^{2} \boldsymbol{k}_{2\perp} \int \frac{d^{2} \boldsymbol{b}_{\perp}}{(2 \pi)^{2}} e^{i\left(\boldsymbol{q}_{\perp}-\boldsymbol{k}_{1\perp}-\boldsymbol{k}_{2\perp}\right) \cdot \boldsymbol{b}_{\perp}} \nonumber\\
&\times f_{1}^{q /p}\left(x_{1}, \boldsymbol{k}_{1\perp}^{2},Q\right) f_{1}^{\bar{q}/p}\left(x_{2}, \boldsymbol{k}_{2 \perp}^{2},Q\right) \nonumber\\
&=\frac{1}{N_{c}} \sum_{q}^{}e_{q}^{2}\int \frac{d^{2}\boldsymbol{b}_{\perp}}{(2\pi)^{2}}e^{i\boldsymbol{q_{\perp}}\cdot \boldsymbol{b}_{\perp}}\tilde{f}_{1}^{q/p}(x_{1},\boldsymbol{b}_{\perp},Q) \tilde{f}_{1}^{\bar{q}/p}(x_{2},\boldsymbol{b}_{\perp},Q),
\label{eq:F-UU}
\end{align}
where the unpolarized distribution functions from each proton in the $b_{\perp}$-space are defined as follows
\begin{align}
\tilde{f}_{1}^{q/p}(x_{1},\boldsymbol{b}_{\perp},Q)=\int d^{2}\boldsymbol{k}_{1\perp}e^{-i\boldsymbol{k}_{1\perp}\cdot \boldsymbol{b}_{\perp}}f_{1}^{q/p} (x_{1},\boldsymbol{k}_{1\perp}^2;Q),
\label{eq:f1-q}
\end{align}
\begin{align}
\tilde{f}_{1}^{\bar{q}/p}(x_{2},\boldsymbol{b}_{\perp},Q)=\int d^{2}\boldsymbol{k}_{2\perp}e^{-i\boldsymbol{k}_{2\perp}\cdot \boldsymbol{b}_{\perp}}f_{1}^{\bar{q}/p} (x_{2},\boldsymbol{k}_{2\perp}^2;Q).
\label{eq:f1-anti-q}
\end{align}

\section{The TMD evolution of distribution functions}
\label{Sec.evolution}

In this section, we systematically review the energy evolution of both unpolarized ($f_1$) and longi-transversity ($h_{1L}^{\perp}$) distributions within the TMD factorization framework. 
Our analysis incorporates recent theoretical developments in TMD resummation techniques and nonperturbative parametrizations.

Based on the TMD factorization theorem, as formulated in various systems (such as CS-81~\cite{Collins:1981uk}, JMY~\cite{Ji:2004wu,Ji:2004xq}, and Collins-11~\cite{Collins:2011zzd}), the distribution function $\widetilde{F} (x,b;\mu ,\zeta_{F} )$ in $b$-space depends on two energy scales: the renormalization scale $\mu$, associated with the corresponding collinear distribution functions, and the energy scale $\zeta _{F}$, which regularizes the light-cone singularity in the operator definition of TMDs. The $\zeta _{F}$ dependence of TMDs is governed by the Collins-Soper (CS) equation~\cite{Collins:1981uk} ($b=|\boldsymbol{b}_{\perp}|$):
\begin{align}
\frac{\partial  \ln\widetilde{F} (x,b;\mu ,\zeta_{F} ) }{\partial \sqrt{\zeta _{F}} }=\widetilde{K} (b;\mu ),
\label{eq:energy scale}
\end{align}
where $\widetilde{K}$ represents the CS evolution kernel, which can be computed perturbatively at small $b$ region. The result up to order $\alpha_{s}$ has the form
\begin{align}
\widetilde{K} (b;\mu )=-\frac{\alpha _{s} C_{F} }{\pi }\left [ \ln(\mu ^{2}b ^{2} )-\ln4+2\gamma _{E} \right ] + \mathcal{O} (\alpha _{s}^{2}),
\label{eq:Evolution nucleus}
\end{align}
where $\gamma _{E}  \approx 0.577$ is the Euler's constant.

The $\mu$ dependence of the TMDs is derived from the renormalization group equation
\begin{align}
\frac{d\widetilde{K} }{d \ln\mu } =-\gamma _{K} (\alpha _{s} (\mu )),
\label{eq:renormalization scale 1}
\end{align}
\begin{align}
\frac{d\ln\widetilde{F} (x,b;\mu ,\zeta _{F}) }{d\ln\mu }  =\gamma _{F} (\alpha _{s} (\mu );\frac{\zeta _{F}^{2} }{\mu^{2}} ),
\label{eq:renormalization scale 2}
\end{align}
where $\gamma _{K}$ and $\gamma _{F}$ are the anomalous dimensions of $\widetilde{K}$ and $\widetilde{F}$, respectively,
\begin{align}
\gamma _{K} =2\frac{\alpha _{s} C_{F} }{\pi }+\mathcal{O} (\alpha _{s}^{2}),
\label{eq:gamma-K}
\end{align}
\begin{align}
\gamma_{F} = \alpha_{s}\frac{C_{F}}{\pi}(\frac{3}{2}-\ln(\frac{\zeta_{F}}{\mu^{2}} ))+\mathcal{O}(\alpha_{s}^{2}).
\label{eq:gamma-F}
\end{align}

By solving Eqs.(\ref{eq:energy scale}), (\ref{eq:renormalization scale 1})~and (\ref{eq:renormalization scale 2}), one can obtain the general solution for the energy dependence of TMDs
\begin{align}
\widetilde{F} (x,b;Q)=\mathcal{F} \times e^{-S(Q,b)} \times\widetilde{F} (x,b;\mu),
\label{eq:F-b-Q}
\end{align}
where $\mathcal{F}$ is the hard scattering factor, and $S(Q,b)$ is the Sudakov-like form factor. 
Here, we have set $\mu=\sqrt{\zeta _{F} }=Q$, allowing us to simplify $\widetilde{F} (x,b;\mu ,\zeta_{F} )$ as $\widetilde{F}(x,b;Q)$.
Eq.(\ref{eq:F-b-Q}) demonstrates that the energy evolution of TMDs from the initial energy $\mu$ to another energy scale $Q$ can be realized through the Sudakov form factor $S(Q,b)$ by the exponential form, $e^{-S(Q,b)}$.

The exponential exp($-S(Q,b)$) for $\widetilde{F}$ can be written as~\cite{Li:2021mmi,Li:2021txj}
\begin{align}
&\text{exp}({-S(Q,b)})=\text{exp}\left \{ \text{ln}\frac{Q}{\mu } \widetilde{K}(b_{\ast  };\mu )  +\int_{\mu _{i} }^{\mu }\frac{d\bar{\mu } }{\bar{\mu} } \times \left [ \gamma _{F}(g(\bar{\mu } ));1)-\text{ln}\frac{Q}{\bar{\mu } }  \gamma _{K}(g(\bar{\mu } ))\right ] \right \}\nonumber\\&
\times \text{exp}\left \{ g_{i/p}(x,b) + g_{K}(b)\text{ln}\frac{Q}{Q_{0} }  \right \}.
\label{eq:Sudakov e-index}
\end{align}
where the first exponential term arises from the perturbative region ($b \ll 1/\Lambda$), containing the CS evolution kernel $\widetilde{K}(b_{\ast}; \mu)$ within the perturbative region, specifically in the small $b$ region where $b\ll 1/\Lambda$, and the anomalous dimensions $\gamma_F$ and $\gamma_K$. 
The second exponential term accounts for nonperturbative effects in the large $b$ region, where $\widetilde{K}(b; \mu)$ cannot be calculated perturbatively. 
Here, $g_{i/p}(x, b)$ parameterizes the intrinsic nonperturbative behavior of parton $i$ within the proton, while $g_K(b)$ describes the nonperturbative behavior of the evolution kernel $\widetilde{K}(b; \mu)$.

To ensure a smooth transition between the perturbative and nonperturbative regions, a boundary parameter $b_{\max} \approx 1~\text{GeV}^{-1}$ is introduced, along with a $b$-dependent function $b_{\ast}(b)$. 
This function is designed to possess the property that $b_{\ast}\approx b$ at small $b$ value and $b_{\ast}\approx b_{\max}$ for large $b$ value. In literatures~\cite{Bacchetta:2017gcc,Collins:1984kg,Collins:2016hqq}, several different expressions for $b_{\ast}(b)$ have been proposed. A commonly used prescription is the Collins-Soper-Sterman (CSS) form~\cite{Collins:1984kg}: 
\begin{align}
b_{\ast}(b)=b/\sqrt{1+b^2/{b^2_{\max}}},~b_{\max}<1/\Lambda_{\text{QCD}}. 
\label{eq:b-ast}
\end{align}

Therefore, the Sudakov form factor $S(Q,b)$ appearing in Eq.(\ref{eq:F-b-Q}) can be decomposed into a perturbative part $S_{P}(Q,b_{\ast})$ and a nonperturbative part $S_{NP}(Q,b)$ as follows
\begin{align}
S(Q,b)=S_{P} (Q,b_{\ast})+S_{NP} (Q,b),
\label{eq:Sudakov}
\end{align}
with the boundary of two parts set by the $b_{\max}$. 
According to the intensive studies in Refs.\cite{Echevarria:2014xaa,Kang:2011mr,Echevarria:2014rua,Echevarria:2012pw,Aybat:2011ge}, 
the perturbative part can be expanded as an $\alpha_{s}/\pi$ series
\begin{align}
S_{P}(Q,b_{\ast})=\int_{\mu^{2}  }^{Q^{2} }\frac{d\bar{\mu } ^{2} }{\bar{\mu } ^{2}} \left [\mathcal{A}(\alpha _{s}(\bar{\mu} ) ) \ln \frac{Q^{2} }{\bar{\mu}^{2}  }+\mathcal{B}(\alpha _{s}(\bar{\mu} ) )  \right ],
\label{eq:SP}
\end{align}
where the coefficients $A$ and $B$ are given by
\begin{align}
\mathcal{A}=\sum_{n=1 }^{\infty } A^{(n)}(\frac{\alpha _{s} }{\pi } )^{n},
\label{eq:coefficient A}
\end{align}
\begin{align}
\mathcal{B}=\sum_{n=1 }^{\infty } B^{(n)}(\frac{\alpha _{s} }{\pi } )^{n}.
\label{eq:coefficient B}
\end{align}

In our calculations, we will take $A^{(n)}$ up to $A^{(2)}$ and $B^{(n)}$ up to $B^{(1)}$ in the accuracy of next-to-leading logarithmic (NLL) order~\cite{Aybat:2011zv,Collins:1984kg,Kang:2011mr,Landry:2002ix,Echevarria:2012pw,Qiu:2000ga}
\begin{align}
&A^{(1)}=C_{F}, \nonumber\\
&A^{(2)}=\frac{C_{F}}{2}\left[C_{A}\left(\frac{67}{18}-\frac{\pi^{2}}{6}\right)-\frac{10}{9} T_{R} n_{f}\right],\nonumber\\
&B^{(1)}=-\frac{3}{2} C_{F},
\label{eq:NLL}
\end{align}
where $C_{F}=\frac{4}{3}$, $C_{A}=3$, $T_{R}=\frac{1}{2}$ and $n_{f}=5$.

The non-perturbative part $S_{NP}$ in Eq.(\ref{eq:Sudakov}) can not be calculated perturbatively, it is usually extracted from experimental data. 
There are several extractions for $S_{NP}$ by different groups~\cite{Aybat:2011zv,Echevarria:2012js,Echevarria:2014xaa,Bacchetta:2017gcc,Collins:1984kg,Collins:2011zzd,Kang:2011mr,Echevarria:2014rua,Landry:2002ix,Davies:1984sp,Ellis:1997sc,Aybat:2011ge,Sun:2014dqm,Collins:2014jpa,Nadolsky:1999kb,Aidala:2014hva,Konychev:2005iy}. In our study we employ the Bacchetta-Delcarro-Pisano-Radici-Signori (BDPRS) parametrization for the unpolarized TMDs~\cite{Bacchetta:2017gcc}:
\begin{align}
S_{NP}^{f_1^{q/p}}=S_{NP}^{f_1^{\bar{q}/p}}=-\frac{1}{2}g_{k}(b)\ln(Q^{2}/Q_{0}^{2})-\ln(\widetilde{f} _{1NP}^{p}(x,b^2)),
\label{eq:BDPRS}
\end{align}
where $g_{k}(b)=-g_{2}b^{2}/2$, following the traditional choice in Refs.~\cite{Landry:2002ix,Nadolsky:1999kb,Konychev:2005iy} with $g_2$ being a free parameter. 
Moreover, the intrinsic nonperturbative part $\widetilde{f} _{1NP}^{p}(x,b^{2})$ of the TMDs can be parameterized in the following form
\begin{align}
\widetilde{f} _{1NP}^{a}(x,b^{2})=\frac{1}{2\pi}e^{-g_{1}\frac{b^{2}}{4}}(1-\frac{\lambda g_{1}^{2}}{1+\lambda g_{1}}\frac{b^{2}}{4}),
\label{eq:f-NP}
\end{align}
with
\begin{align}
g_{1}(x)=N_{1}\frac{(1-x)^{\alpha }x^{\sigma}}{(1-\widehat{x}^{\alpha } )\widehat{x}^{\sigma}},
\label{eq:g1}
\end{align}
Here $\widehat{x}$ is fixed as $\widehat{x}=0.1$. and  $\alpha$, $\sigma$ and $N_{1}\equiv g_{1}(\widehat{x})$ are free parameters fitted to the available data from SIDIS, Drell–Yan, and $W/Z$ boson production processes. 
In Ref.\cite{Bacchetta:2017gcc}, a new $b_{\ast}(b)$ prescription different from Eq.(\ref{eq:b-ast}) was proposed as
\begin{align}
b_{\ast}(b)=b_{\max}\left(\frac{1-e^{-b^{4}/b_{\max}^{4}  } }{1-e^{-b^{4}/b_{\min}^{4}  } } \right )^{1/4}. 
\label{eq:b ast}
\end{align}
Again, $b_{\max}$ serves as the boundary separating the nonperturbative and perturbative regions in the $b$ space. It is assigned a fixed value of $b_{\max}=2e^{-\gamma_{E}}~\text{GeV}^{-1}\approx 1.123~\text{GeV}^{-1}$. Moreover, the authors in Ref.~\cite{Bacchetta:2017gcc} also choose to saturate $b_{\ast}$ at the minimum value, where $b_{\min}\propto2e^{-\gamma_{E}}/Q$.

Besides the Sudakov form factor in Eq.(\ref{eq:F-b-Q}), another crucial element within Eq.(\ref{eq:F-b-Q}) is the TMDPDFs at a fixed scale $\mu$. In the small $b$ region, $\widetilde{F} (x,b;\mu)$ at $\mu$ can be expressed as a convolution of the perturbatively calculable coefficients $C$ and the corresponding collinear TMDs~\cite{Collins:1981uk,Bacchetta:2013pqa}
\begin{align}
\widetilde{F}(x,b;\mu)=\sum_{i} C_{q\gets i} \otimes F_{i/H}(x,\mu),
\label{eq:F-b-mu}
\end{align}
where $\sum_{i}$ runs over both quark and antiquark flavors, and $\otimes$ denotes the convolution in the longitudinal momentum fraction $x$
\begin{align}
C_{q\gets i} \otimes F_{i/H}(x,\mu) \equiv \int_{x}^{1}\frac{d\xi }{\xi}C_{q\gets i}(x/\xi,b;\mu)F_{i/H}(\xi,\mu).
\label{eq:x convolution}
\end{align}
Here, $F_{i/H}(\xi,\mu)$ is the corresponding collinear TMDs of flavor $i$ in hadron $H$ at the scale $\mu$, which could be a dynamic scale related to $b_{\ast}$ by $\mu=c_{0}/b_{\ast}$, with $c_{0}=2e^{-\gamma_{E}}$~\cite{Collins:1981uk}.

Therefore, with the TMD evolution, the scale-dependent TMDs $\widetilde{F}_{q/H}(x,b;Q)$ can be expressed as 
\begin{align}
\widetilde{F}_{q/H}(x,b;Q)=e^{-\frac{1}{2}S_{P} (Q,b_{\ast })-S_{NP} ^{F_{q/H}}(Q,b)} \mathcal{F}(\alpha _{s}(Q)) \sum_{i}^{} C_{q\gets i} ^{F}\otimes F_{i/H}(x,\mu),
\label{eq:TMD-F}
\end{align}
where the factor $\frac{1}{2}$ in front of $S_{P}$ comes from the fact that $S_{P}$ of quarks and antiquarks satisfies the following relation~\cite{Prokudin:2015ysa}
\begin{align}
S_{P}^{q}(Q,b_{\ast} ) =S_{P}^{\bar{q}}(Q,b_{\ast} )=S(Q,b_{\ast} )/2.
\label{eq:Spp}
\end{align}
In addition, the hard coefficients $C$, $\mathcal{F}$ for $f_{1}$ and $h_{1L}^{\perp}$ have been calculated up to next-to-leading order (NLO). 
Nevertheless, only the first term of the $h_{1L}^{\perp}$ result in Eq. (60) of Ref.~\cite{Zhou:2009jm} (specifically, the $\tilde{h}(x)$ term)  dominates. 
In this study, there is no need to consider the contribution of $\tilde{T} _{F}^{(\sigma)} $ because it is beyond the Wandzura-Wilczek(WW)-approximation. 
For consistency, here we adopt the leading-order results for the hard coefficients $C$ and $\mathcal{F}$ for $f_{1}$ and  $h_{1L}^{\perp}$, where $\mathcal{F}=1$ and $C_{q\gets i} ^{0}=\delta_{iq}\delta(1-x)$.

Using the above choices, the expression for the unpolarized TMD of the proton $\widetilde{f} _{1}^{q/p}$ in $b$-space reduces to~\cite{Bacchetta:2017gcc}
\begin{align}
\widetilde{f} _{1}^{q/p}(x,b;Q)=e^{-\frac{1}{2} S_{P}(Q,b_{\ast } ) -S_{NP}^{f_{1}^{q/p}} }\widetilde{f} _{1}^{q/p}(x,\mu).
\label{eq:f1-b-Q}
\end{align}
By performing the Fourier Transformation, we can convert the function $\widetilde{f} _{1}(x,b;Q) $ into the transverse momentum space
\begin{align}
f_{1}^{q/p}(x,k_{\perp } ;Q)=\int_{0}^{\infty }  \frac{dbb}{2\pi }  J_{0} (|\boldsymbol{k}_{\perp}| b)  e^{-\frac{1}{2} S_{P}(Q,b_{\ast } ) -S_{NP}^{f_{1}^{q/p}} }f _{1,q/p}(x,\mu),
\label{eq:f1-k-Q}
\end{align}
where $J_{0} $ is the Bessel function of the first kind.

According to Eqs.(\ref{eq:F-b-mu}) and (\ref{eq:x convolution}), in the small $b$ region, one can express the longi-transversity of the incoming protons at scale $\mu$ in terms of the perturbatively calculable coefficients and the corresponding collinear correlation function as follows~\cite{Boer:2011xd,Bastami:2018xqd}
\begin{align}
\tilde{h}_{1L}^{\perp(\alpha)~q/p}(x,b;\mu)=ib^{\alpha}Mh_{1L}^{\perp(1)}(x,\mu).
\label{eq:h1L-perp}
\end{align}
Here, the superscript (1) denotes the first transverse moment of the longi-transversity $h_{1L}^{\perp}$, which is defined as~\cite{Boer:2011xd,Bastami:2018xqd}
\begin{align}
h_{1L}^{\perp~(1)}(x,\mu)=\int d^{2}\boldsymbol{k}_{\perp}\frac{\boldsymbol{k}_{\perp}^{2}}{2M^{2}}h_{1L}^{\perp}(x,\boldsymbol{k}_{\perp}^2;\mu).
\label{eq:h1L-perp(1)}
\end{align}

Regarding the nonperturbative part of the Sudakov form factor associated with the longi-transversity, the information still remains unknown. In practical calculations, we assume that it is the same as that for the unpolarized distribution function $S_{NP}^{f_{1}^{q/p}}$. Thus, we can obtain the longi-transversity in $b$-space as
\begin{align}
\tilde{h}_{1L}^{\perp(\alpha)~q/p}(x,b;Q)=ib^{\alpha}Me^{-\frac{1}{2} S_{P}(Q,b_{\ast } ) -S_{NP}^{f_{1}^{q/p}}}h_{1L}^{\perp(1)}(x,\mu).
\label{eq:h1L-perp-alpha}
\end{align}

After performing the Fourier transformation, the longi-transversity in the transverse momentum space is given by
\begin{align}
\frac{{k}_{\perp}^{\alpha}}{M}h_{1L}^{\perp q/p} \left ( x,{k}_{\perp};Q \right )
=M\int_{0}^{\infty}\frac{dbb^{2}}{2\pi}J_{1}(\left |\boldsymbol{k}_{\perp}  \right |b )e^{-\frac{1}{2}S_{P}(Q,b_{\ast })-S_{NP}^{f_{1}}} \times{h}_{1L}^{\perp (1)}(x,\mu) .
\label{eq:h1L-perp~}
\end{align}

With all ingredients described above, Eqs.(\ref{eq:F-LL}) and (\ref{eq:F-UU}) can be rewritten in the following forms
\begin{align}
F_{LL}^{cos2\phi }
=\frac{1}{N_{c}} \sum_{q} e_{q}^{2} \int_{0}^{\infty} \frac{dbb^{3}}{2 \pi} J_{2}\left(|\boldsymbol{q}_{\perp}|b\right)M^2 h^{\perp(1),q/p}_{1L}\left(x_{1}, \mu\right) h^{\perp(1), \bar{q}/ p}_{1L}\left(x_{2}, \mu\right) 
\times e^{-\left(S_{\mathrm{NP}}^{f_{1}, q / p}+S_{\mathrm{NP}}^{f_{1}, \bar{q} / p}+S_{\mathrm{P}}\right)},&
\label{eq:F-LL-final}
\end{align}
\begin{align}
F_{UU}^{1}
=\frac{1}{N_{c}} \sum_{q} e_{q}^{2} \int_{0}^{\infty} \frac{dbb}{2 \pi} J_{0}\left(|\boldsymbol{q}_{\perp}|b\right) f_1^{q/p}(x_1, \mu) f_1^{\bar{q} / p}(x_2, \mu) 
\times e^{-\left(S_{\mathrm{NP}}^{f_{1}, q / p}+S_{\mathrm{NP}}^{f_{1}, \bar{q} / p}+S_{\mathrm{P}}\right)}.&
\label{eq:F-UU-final}
\end{align}

\section{Numerical calculation}
\label{Sec.numerical}

In this section, we present a comprehensive numerical analysis of the $\cos2\phi$ azimuthal asymmetry in double-longitudinally polarized proton-proton Drell-Yan collisions, utilizing the theoretical framework established in previous sections. Our calculations are performed for the kinematic regimes accessible at both RHIC and NICA, with particular emphasis on the role of sea quark contributions.

The calculation requires the collinear functions $h^{\perp(1)}_{1L}(x,\mu)$ and $f_1(x,\mu)$ as inputs to Eqs.~(\ref{eq:F-LL-final}) and (\ref{eq:F-UU-final}). For the unpolarized distribution $f_1(x,\mu)$, we employ the NLO CT10 parametrization (central PDF set) from Ref.~\cite{Lai:2010vv}. 
For the longitudinally polarized transversity distribution $h^{\perp(1)}_{1L}(x,\mu)$, we adopt the Wandzura-Wilczek approximation~\cite{Jaffe:1991ra,Bastami:2018xqd}:
\begin{align}
h_{1L}^{\perp(1)}(x)\overset{WW-type}{\approx}-x^{2}\int_{x}^{1} \frac{dy}{y^{2}} h_{1}(y),
\label{eq:h1y}
\end{align}
where $h_{1}$ is the transversity distribution. For the velance quark component of the transversity distribution, we select the parametrization extracted from Ref.~\cite{Kang:2015msa} which employed the TMD evolution formalism. 
At the initial scale $Q_0=\sqrt{2.4}~\text{GeV}$, it is given by:
\begin{align}
h_{1}^{q}(x,Q_{0})=N_{q}^{h}x^{a_{q}}(1-x)^{b_{q}}\frac{(a_{q}+b_{q})^{a_{q}+b_{q}}}{a_{q}^{a_{q}}b_{q}^{b_{q}}}\times \frac{1}{2}\left (f_{1}^{q}(x,Q_{0})+g_{1}^{q}(x,Q_{0})\right ).
\label{eq:h1y1}
\end{align}
Here $g_{1}^{q}$ is the helicity distribution function~\cite{deFlorian:2009vb}, and $f_{1}^{q}$ is the unpolarized distribution function. For the sea quark component of the transversity distribution, we make the following assumption 
\begin{align}
h_{1}^{sea}(x,Q_{0})=N\frac{1}{2}\left (f_{1}^{q}(x,Q_{0})+g_{1}^{q}(x,Q_{0})\right ),
\label{eq:h1y1}
\end{align}
where $N$ represents the proportion of the contribution of sea quarks in the proton. 
In this calculation, we take into account three scenarios for the contribution ratio of sea quarks in the proton: $N=0$, $N=0.5$ and $N=1$.

We employ the QCDNUM package\cite{Botje:2010ay}  to carry out the evolution of $f_1$ from the initial scale $Q_0=\sqrt{2.4}~\text{GeV}$ to another energy. Concerning the energy evolution of $h^{\perp(1)}_{1L}$, the effect has been studies in Ref~\cite{Zhou:2008mz,Kang:2015msa,Wang:2018pmx,Kang:2008ey}. For simplicity, we only consider the homogenous terms in the evolution kernel
\begin{align}
P_{qq}^{h}=C_{F}\left [ \frac{2x}{(1-x)_{+}} +2\delta (1-x)\right ] -\frac{C_{F}}{2}\frac{2x}{(1-x)}.
\label{eq:37}
\end{align}

Drell-Yan experiment at RHIC is proposed to use two proton beams colliding at either $\sqrt{s}=200~\text{GeV}$ or $500~\text{GeV}$~\cite{Bunce:2000uv}. 
In this study, we estimate the $A_{LL}^{\cos2\phi}$ asymmetry at the kinematical regions of RHIC~\cite{Lu:2011cw}:
\begin{center}
$4~\text{GeV}<~\text{Q}~< 9~\text{GeV}$,~~~~$0<q_{T}<1~\text{GeV}$,~~~~$1<y<2$.   
\end{center}
We also estimate the $A_{LL}^{\cos2\phi}$ asymmetry at the kinematical regions of NICA~\cite{Lu:2011cw}: 
\begin{center} 
$\sqrt{s}=27~\text{GeV}$,~~~~~~~~~~~~$4~\text{GeV}<~Q~<9~\text{GeV}$,\\
$0~<q_{T}<1~\text{GeV}$,~~~~~~~~~~~~$0.1~<x<~0.8 $ .  
\end{center}.

\begin{figure}[htbp]
\centering
\includegraphics[width=0.328\columnwidth]{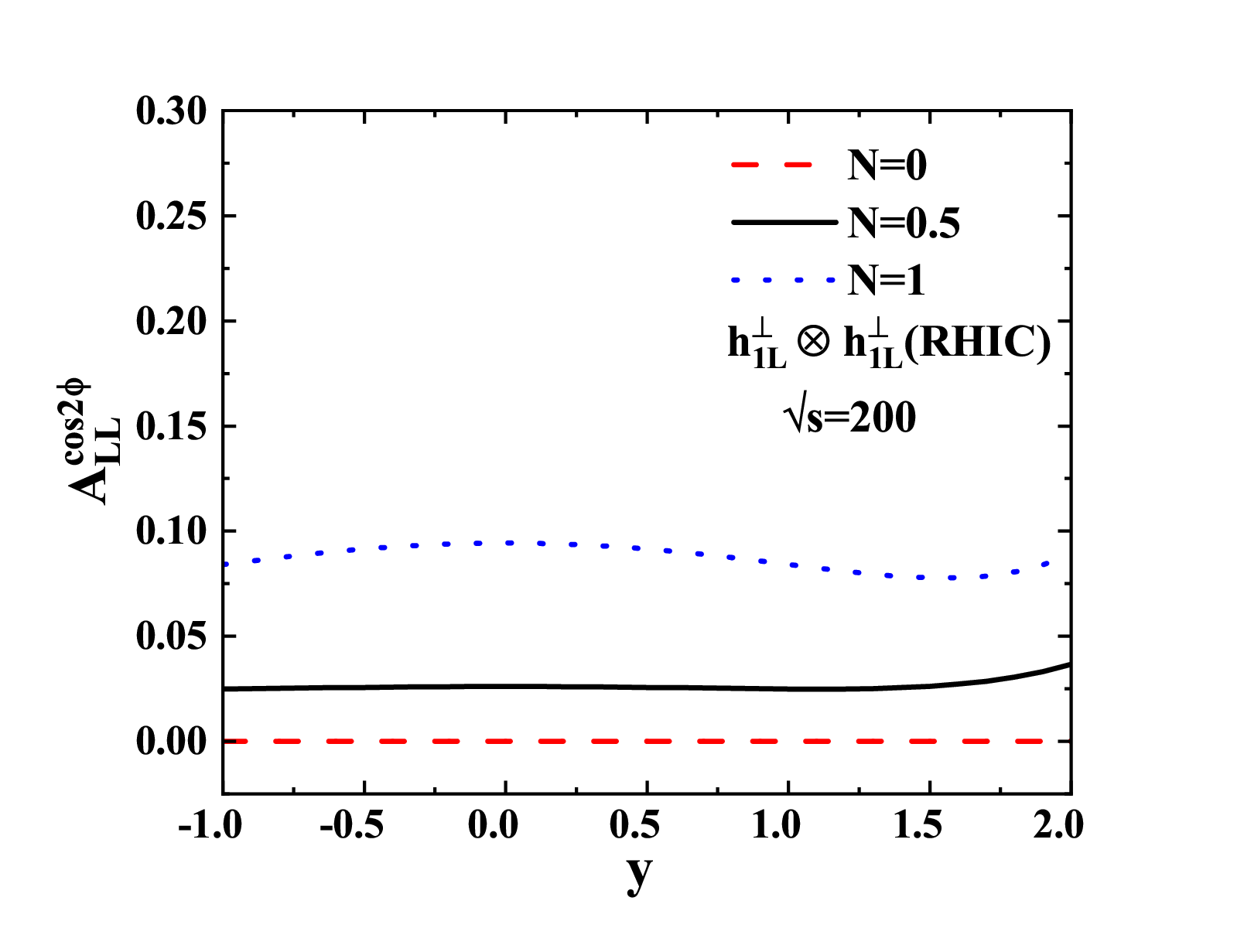}
\includegraphics[width=0.328\columnwidth]{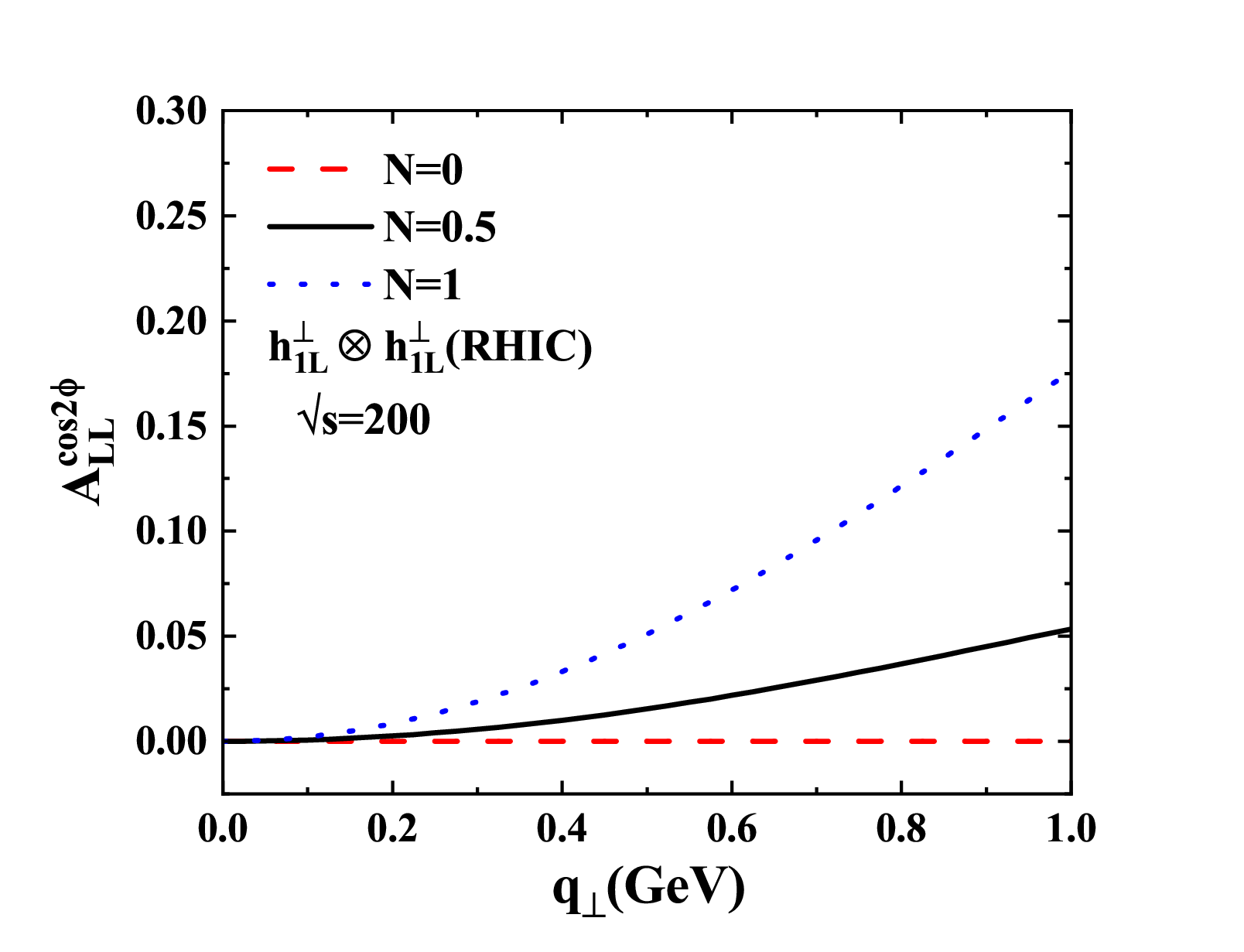}
\includegraphics[width=0.328\columnwidth]{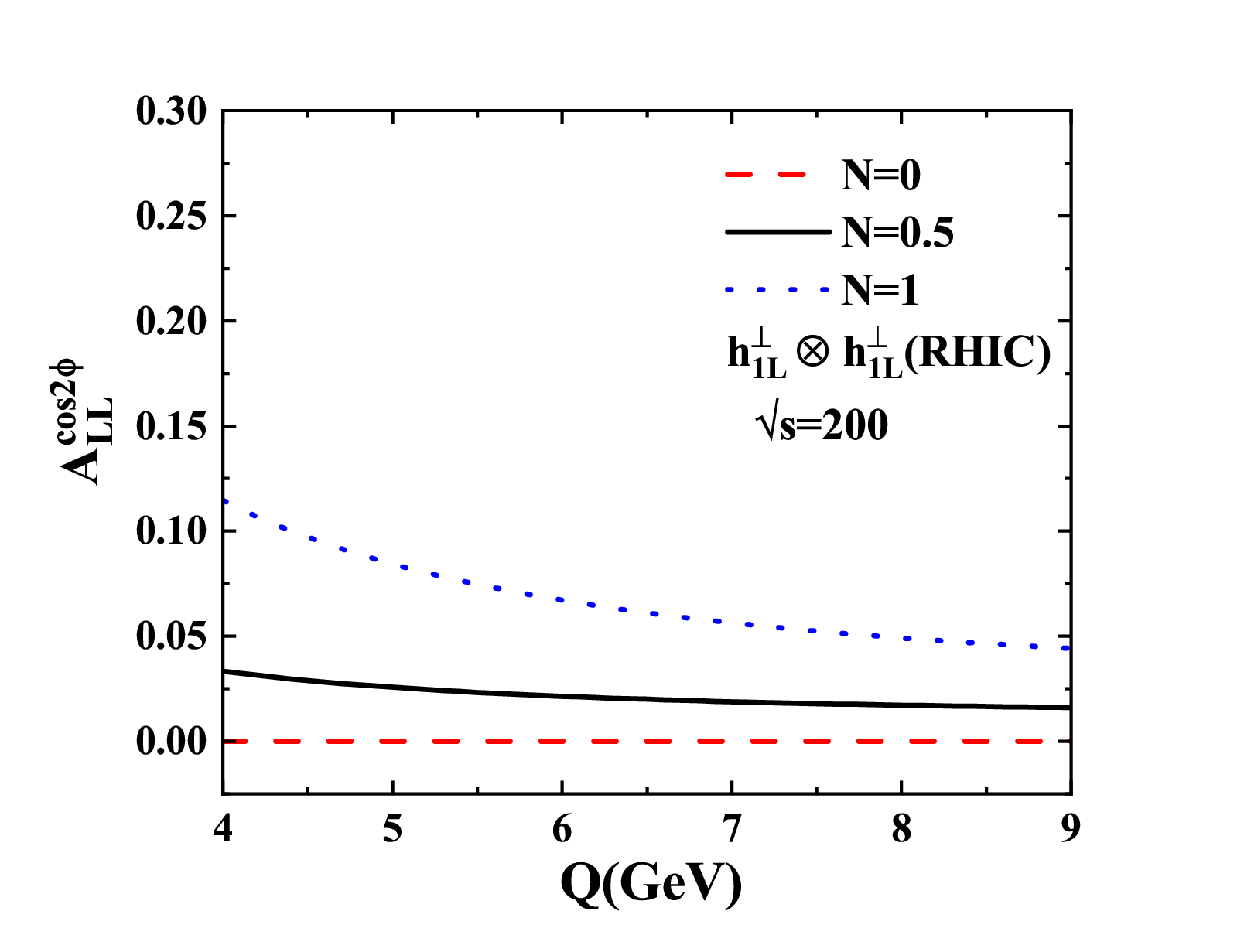}
\caption{ The asymmetry $A_{LL}^{cos(2\phi)}$ in the double-longitudinally polarized $pp$ Drell-Yan process, at the kinematic range of RHIC with $\sqrt{s}=200~ \text{GeV}$, as functions of $y$~(left panel), $q_{\perp}$~(middle panel) and $Q$~(right panel). The dashed line, solid line and dotted line depict the asymmetry for $N=0, 0.5, 1$, respectively.
}
\label{fig1}
\end{figure}

\begin{figure}[htbp]
\centering    \includegraphics[width=0.328\columnwidth]{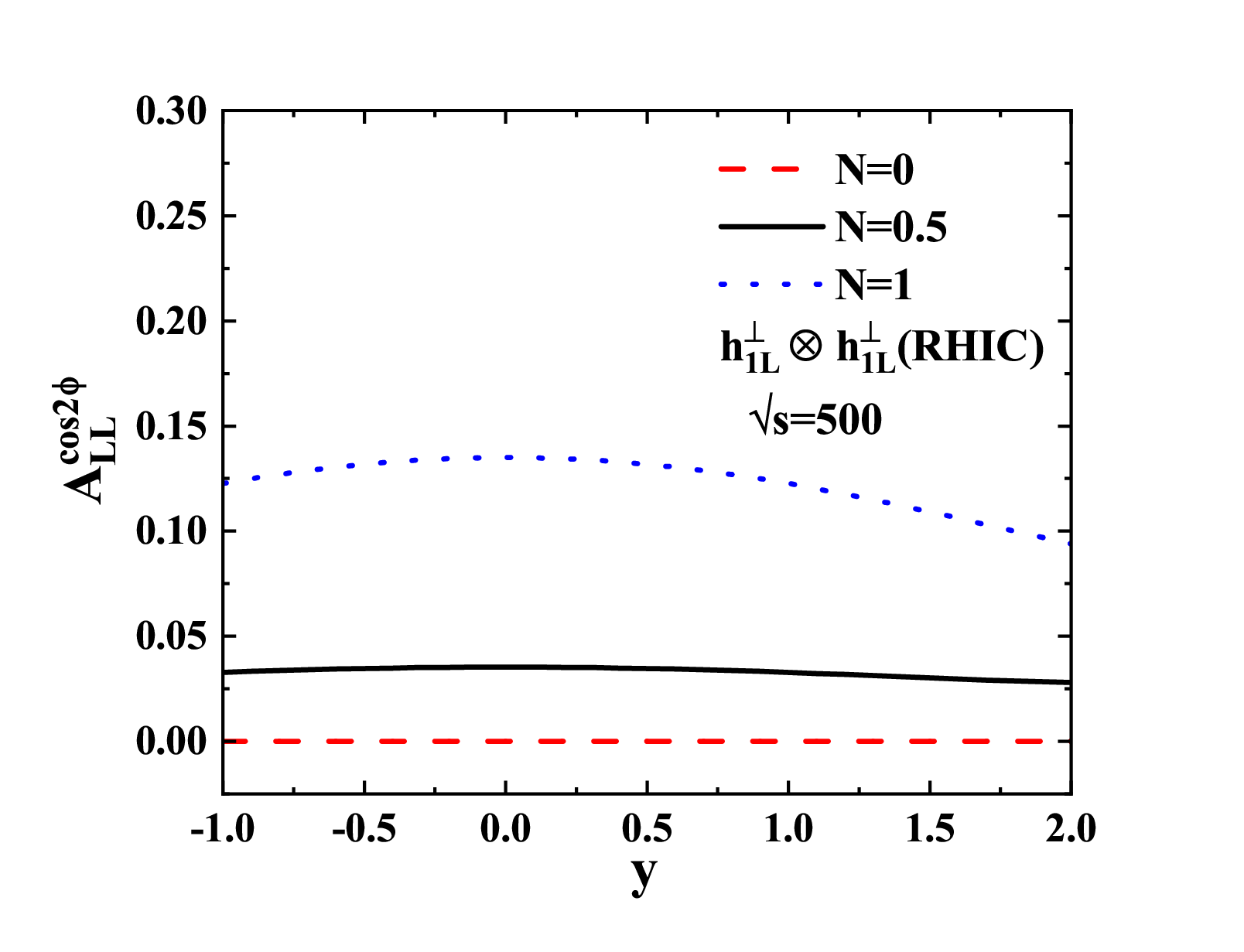}
\includegraphics[width=0.328\columnwidth]{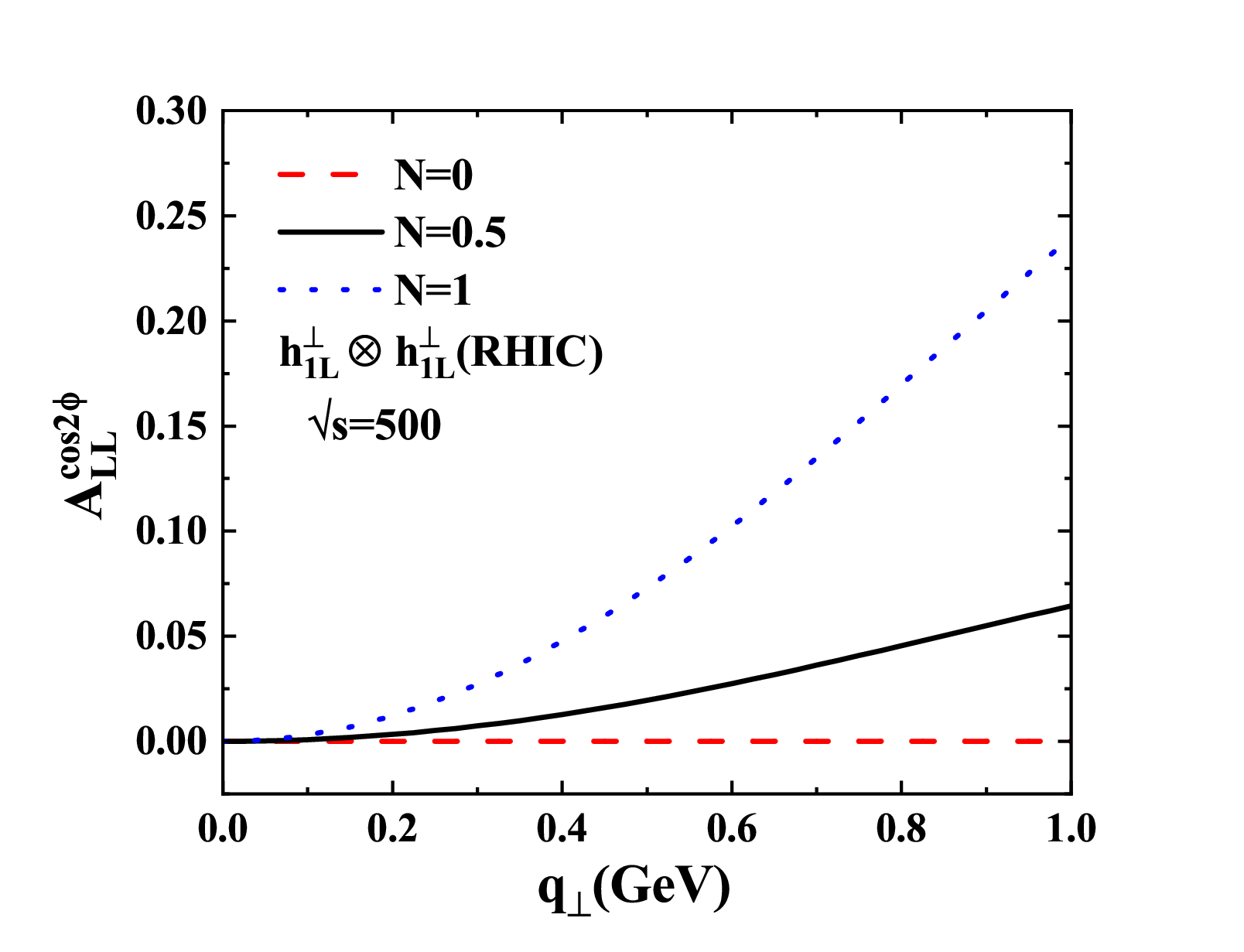}
\includegraphics[width=0.328\columnwidth]{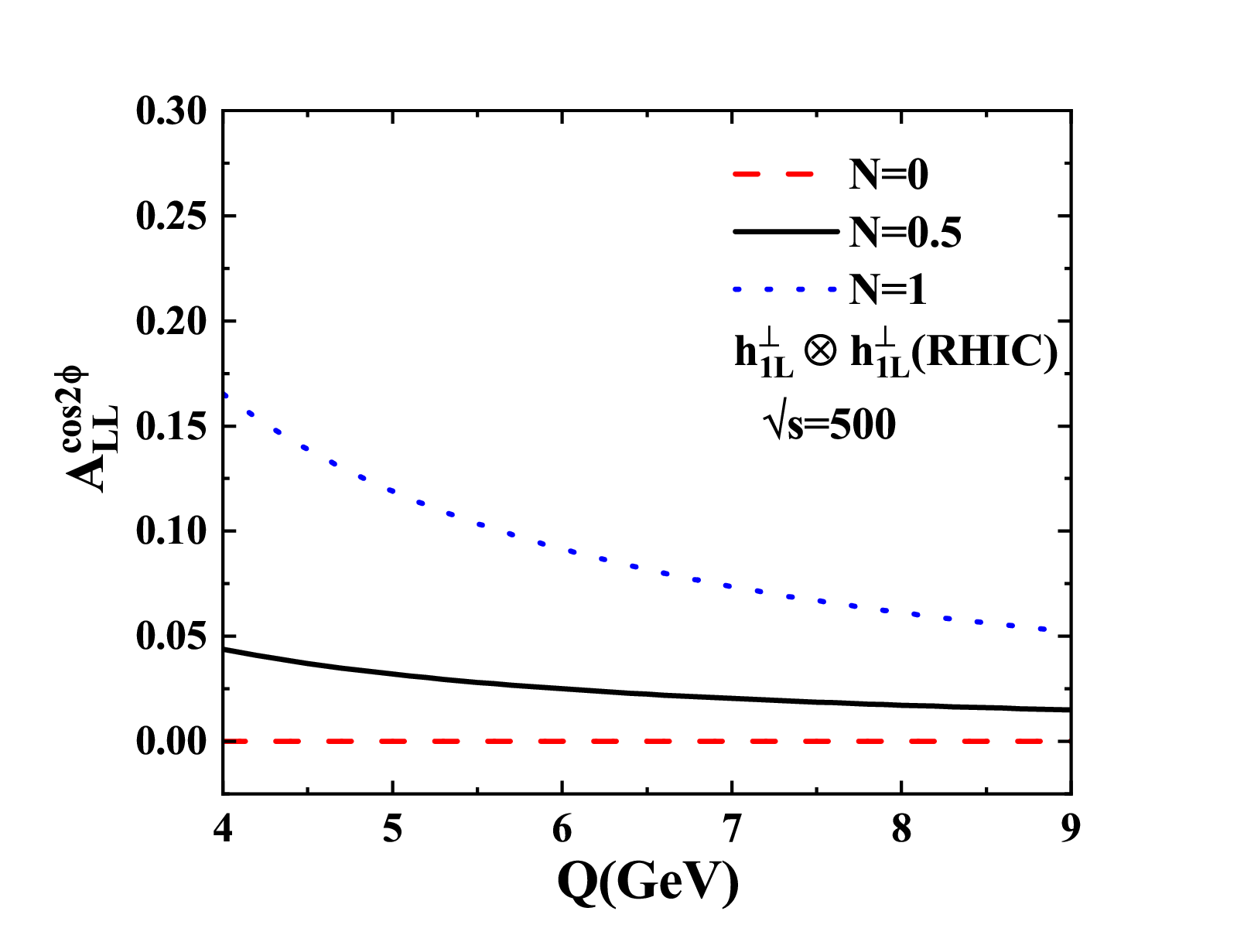}
\caption{Similar to Fig.~\ref{fig1}, but for the asymmetry at the kinematic range of RHIC with $\sqrt{s}=500~\text{GeV}$.}
\label{fig2}
\end{figure}

Figures~\ref{fig1} and \ref{fig2} present our numerical predictions for the $\cos2\phi$ azimuthal asymmetry in $pp$ Drell-Yan collisions at RHIC kinematics. The left, middle, and right panels display the rapidity ($y$), transverse momentum ($q_{\perp}$), and invariant mass ($Q$) dependencies of the asymmetry, respectively. Figure~\ref{fig1} corresponds to $\sqrt{s}=200~\text{GeV}$, while Fig.~\ref{fig2} shows results for $\sqrt{s}=500~\text{GeV}$. The dashed, solid, and dotted lines represent the asymmetry predictions for sea quark contribution parameters $N=0$, $N=0.5$, and $N=1$, respectively.


As shown in Fig.~\ref{fig1}, in all the cases except $N=0$ (where the asymmetry is zero), the $\cos2\phi$ azimuthal asymmetry in the longitudinally polarized proton-proton Drell-Yan process from our calculation is positive. Our estimates also show that the asymmetry changes slightly with the change of $y$. For $N=0.5$, the magnitude of the $y$-dependent asymmetries is in the range of $2\%$ to $5\%$, while for $N=1$, it is in the range of $8\%$ to $10\%$. 
For the $q_\perp$ asymmetry, we find a strong growth with $q_\perp$ for $N=1$, reaching $17.5\%$ at $q_\perp\approx1~\text{GeV}$. Moderate increase of the asymmetry with increasing $q_\perp$ is found for $N=0.5$, peaking at $5\%$ . 
For the $Q$-dependent asymmetry, $A_{LL}^{\cos2\phi}$ moderately decreases from $12\%$ to $5\%$ as $Q$ increases from $4$ to $9~\text{GeV}$ when $N=1$. Weaker $Q$ dependence is observed for $N=0.5$ ($4\%$-$2\%$). The same trendency are observed in Fig. 2, but with larger magnitudes (not more than $25\%$).

\begin{figure}[htbp]
\centering
\includegraphics[width=0.328\columnwidth]{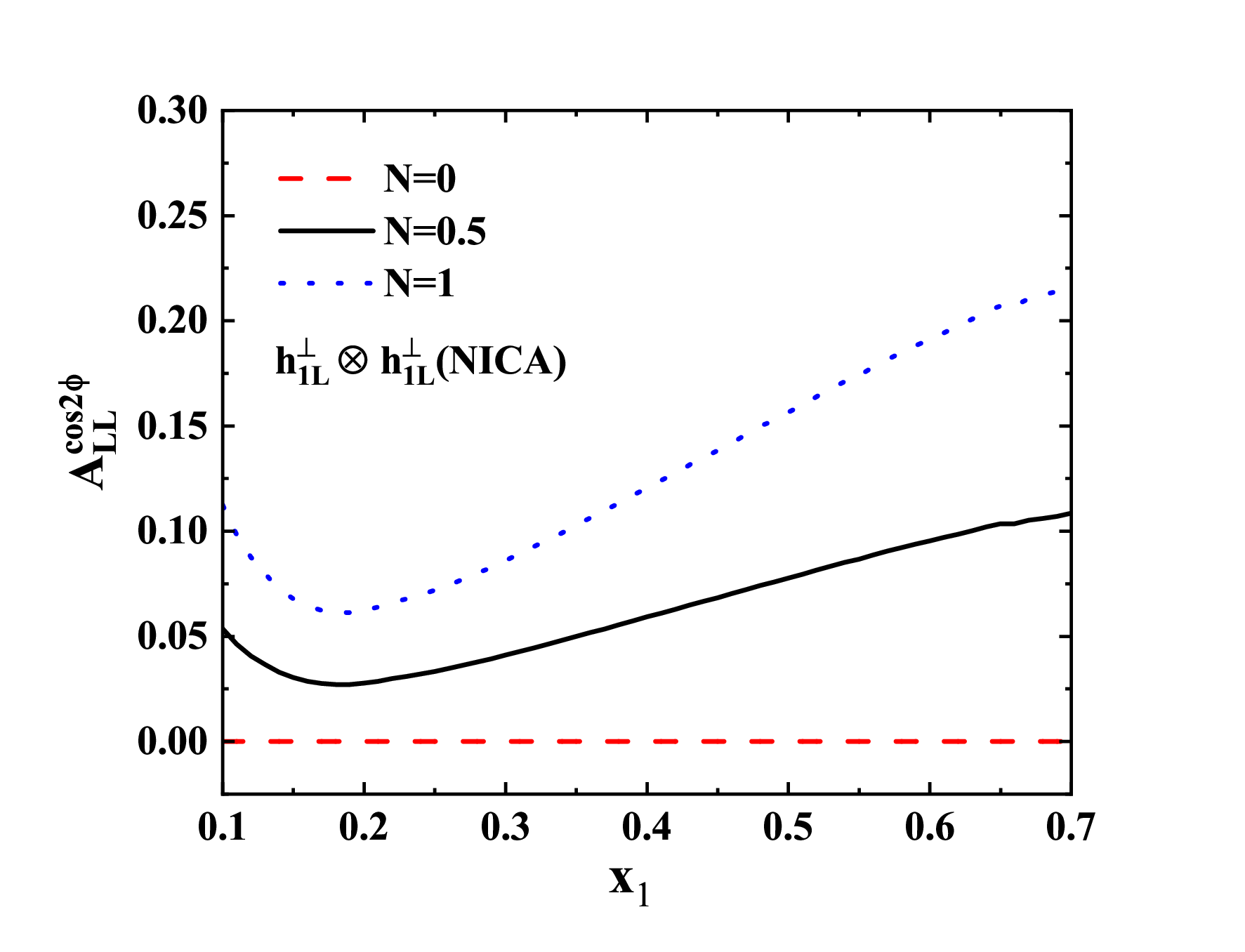}
\includegraphics[width=0.328\columnwidth]{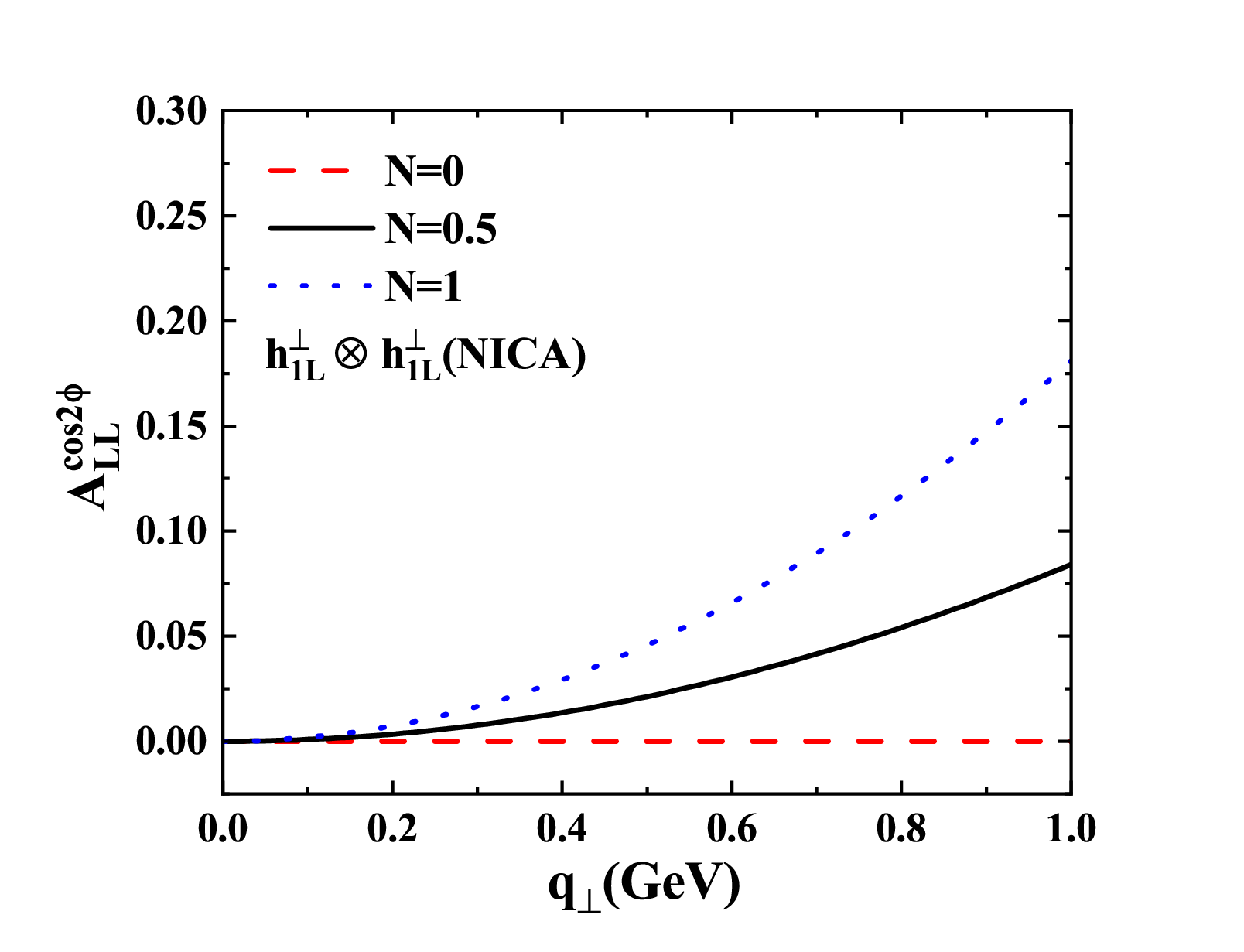}
\includegraphics[width=0.328\columnwidth]{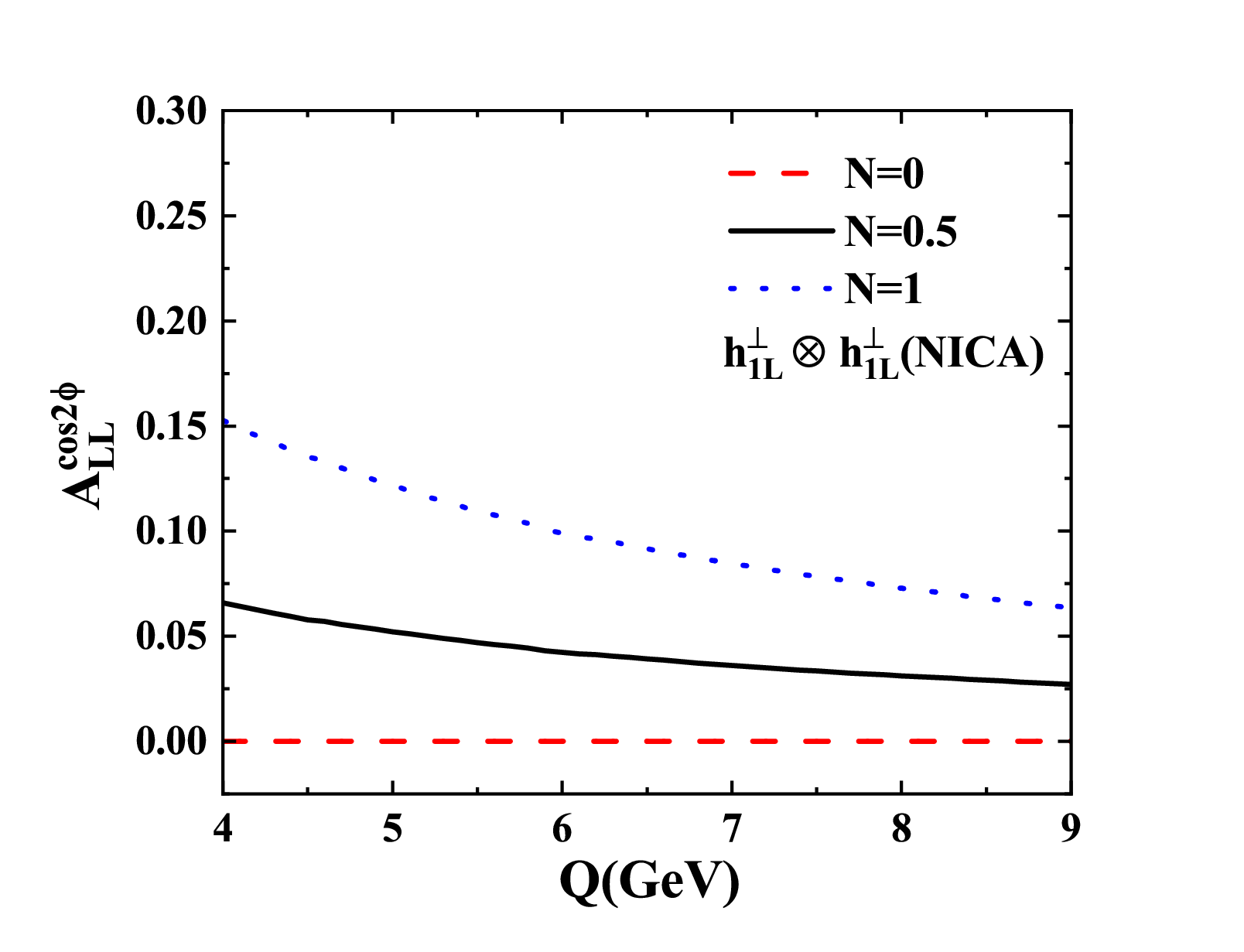}
\caption{Similar to Fig.~\ref{fig1}, but the asymmetry at the kinematic range of NICA.}
\label{fig3}
\end{figure}

The NICA predictions ($\sqrt{s}=27~\text{GeV}$) in Fig~\ref{fig3} exhibit similar qualitative behavior to RHIC results but with different quantitative features. For the Bjorken-$x$ dependence, we observe non-monotonic behavior in the asymmetry for $N=1$, with a maximum value reaching $22\%$. The asymmetry initially decreases in the range $0.1<x<0.2$, followed by a gradual increase up to $x\approx0.7$. This behavior reflects the complex interplay between valence and sea quark contributions in the longitudinally polarized proton.
In terms of TMD dependence, the $q_\perp$ behavior at NICA shows similar trends to RHIC but with reduced magnitudes. The maximum asymmetry reaches $17\%$ for $N=1$, demonstrating the sensitivity of the measurement to the transverse momentum distribution of quarks within the proton.
The $Q$-dependence at NICA exhibits comparable behavior to the RHIC results, with the asymmetry showing a gradual decrease as $Q$ increases. The peak asymmetry reaches $15\%$ for $N=1$, consistent with expectations from TMD factorization. These results highlight the importance of measuring the asymmetry across different kinematic regimes to fully constrain the sea quark contributions and their dependence on the hard scale $Q$. 

We also provide an estimate of the Drell-Yan event rate at NICA. The projected luminosity at NICA is $\mathcal{L}= 10^{32}\text{cm}^{-2}\text{s}^{-1}$~\cite{SPDproto:2021hnm}, with an annual operation time of $3\times10^{7}$ seconds, the integrated luminosity can be obtained as
\begin{align}
\mathcal{L_{\text{int}}}=\mathcal{L} \times t \approx 3 \times 10^{39}\,\mathrm{cm}^{-2}.
\end{align} 
The unpolarized cross section $\sigma^{DY}$ calculated in our framework (Eq.~(\ref{eq:cross section})) is initially in natural units ($\mathrm{GeV}^{-2}$). Converting it to experimental units $\mathrm{cm}^2$ via the relation $1\,\mathrm{GeV}^{-2} \approx 3.894 \times 10^{-28}\,\mathrm{cm}^2$, we obtain $\sigma^{\text{DY}} \approx 5 \times 10^{-37}\,\mathrm{cm}^2$. The expected number of events per year can be estimated as
\begin{align}
N_{\text{events}}=\sigma^{\text{DY}} \times \mathcal{L}_{\text{int}} \approx 1600.
\label{eq:N_events}
\end{align} 
This yield suggests that the proposed asymmetry $A_{LL}^{\cos2\phi}$ could be experimentally accessible at NICA.

In summary, as demonstrated in Figs.~\ref{fig1}-\ref{fig3}, the $\cos2\phi$ azimuthal asymmetries in longitudinally polarized proton-proton Drell-Yan collisions are consistently positive for all cases except $N=0$, where the asymmetry vanishes. Specifically, the asymmetry reaches a maximum of $25\%$ for $N=1$ and up to $10\%$ for $N=0.5$, highlighting a strong dependence on the longitudinally polarized transversity distribution of sea quarks. These results underscore the potential of future precision measurements at RHIC and NICA to significantly enhance our understanding of nucleon structure and QCD dynamics. Moreover, the doubly polarized $p^{\to}p^{\to}$ Drell-Yan process emerges as a powerful probe for elucidating the role of sea quarks in the proton's internal structure, providing crucial insights into the nonperturbative regime of strong interactions.

\section{Conclusion}
\label{Sec.conclusion}

In this study, we employed the TMD factorization framework to explore the $\cos2\phi$ azimuthal asymmetry in double-longitudinally polarized proton-proton Drell-Yan collisions at the kinematic regimes relevant to the RHIC and NICA. This asymmetry originates from the coupling of the longi-transversity distributions $h_{1L}^{\perp}$ from both proton beams. 
To incorporate the scale evolution of TMDs, we introduced the Sudakov form factor for $h_{1L}^{\perp}$, which consists of perturbative and nonperturbative components. For the perturbative part, we adopted the result up to the NLL accuracy, while the nonperturbative part was modeled by the BDPRS parametrization. The hard coefficients associated with the corresponding collinear functions were retained at leading-order precision.

For  the longi-transversity distribution  of the proton, we adopted the WW approximation. Using this framework, we computed the $\cos2\phi$ azimuthal asymmetry in the $p^{\to}p^{\to}$ Drell-Yan process at RHIC and NICA. 
Our results revealed a sensitivity of the asymmetry to the sea quark contributions, parameterized by $N$. Specifically, for $N = 1$, the asymmetry reaches a maximum value of approximately $25\%$; for $N = 0.5$, it decreases to around $10\%$.
This indicates that the sea quark contribution plays an important role in shaping the observed effect.

Our analysis demonstrates that the study of $\cos 2\phi$ asymmetry  can provide quantitative constraints on the longi-transversity distribution within the proton. 
Future high-precision measurements of this asymmetry at RHIC and NICA could provide valuable information about valence and sea quark distributions, further enhancing our understanding of proton structure.

\end{document}